\documentclass[twocolumn,prl,aps]{revtex4}
\usepackage{graphicx}

\newcommand{\be}{\begin{equation}}
\newcommand{\ee}{\end{equation}}
\newcommand{\bea}{\begin{eqnarray}}
\newcommand{\eea}{\end{eqnarray}}
\newcommand{\HH}{{\cal H}}

\newcommand{\la}{\langle}
\newcommand{\ra}{\rangle}
\newcommand{\lb}{\left[}
\newcommand{\rb}{\right]}
\newcommand{\lp}{\left(}
\newcommand{\rp}{\right)}
\renewcommand{\vec}[1]{{\bf #1}}

\begin{document}
\title{Synchronization in the BCS Pairing Dynamics as a Critical Phenomenon}
\author{R.\,A. Barankov$^1$ and L.\,S. Levitov$^2$}
\affiliation{
$^1$
Department of Physics, University of Illinois at Urbana
Champaign,  1110 W. Green St, Urbana, IL 61801,
\\$^2$Department of Physics,
Massachusetts Institute of Technology, 77 Massachusetts Ave, Cambridge, MA 02139}

\begin{abstract}
Fermi gas with time-dependent pairing interaction hosts several
different dynamical states. Coupling
between the collective BCS pairing mode and individual Cooper pair states
can make the latter either synchronize or dephase.
We describe transition from phase-locked undamped oscillations
to Landau-damped dephased oscillations in the collisionless,
dissipationless regime as a function of coupling strength.
In the dephased regime, we find a second transition at
which the long-time asymptotic pairing amplitude vanishes.
Using a combination of
numerical and analytical methods we establish 
a continuous (type II)
character of both transitions.
\end{abstract} \pacs{} \keywords{}

\maketitle

\begin{figure}[t]
\includegraphics[width=3.4in]{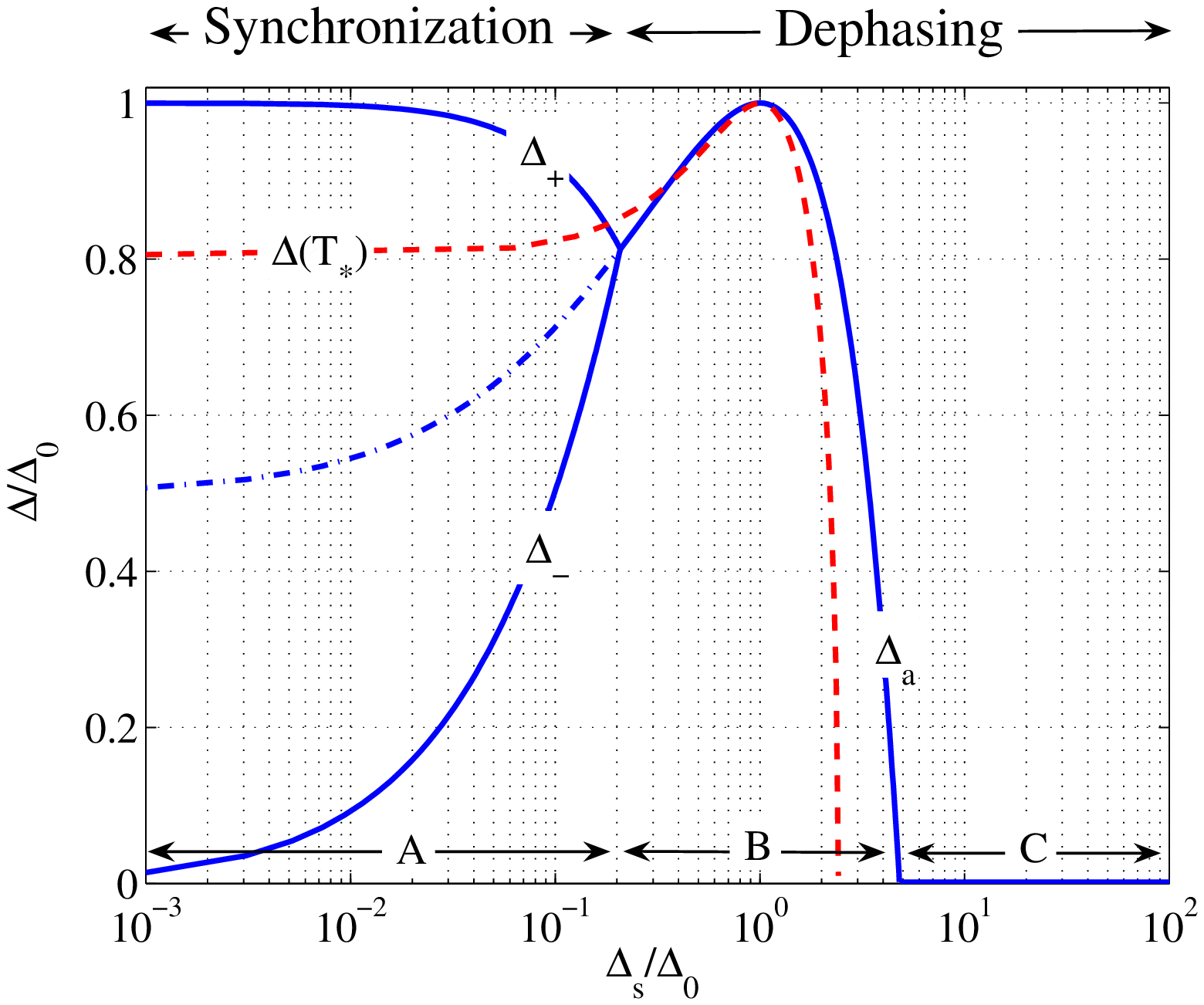}
\vspace{-0.25cm} \caption[]{Three regimes
of the pairing
dynamics {\it vs.} the initial gap value $\Delta_s$.
In synchronized phase ($A$),
$\Delta_s<\Delta_{AB}$,
the pairing amplitude oscillates between
$\Delta_-$ and $\Delta_+$.
In the dephased regime ($B$, $C$),
the pairing amplitude saturates to a constant value, $\Delta_a$,
when
$\Delta_{AB}\le\Delta_s<\Delta_{BC}$,
and decreases to zero at
$\Delta_s\ge \Delta_{BC}$.
Dashed line: The stationary gap value  $\Delta(T_*)$
reached in a closed system after equilibration.
}
%
\label{fig:dyn_regimes}
\end{figure}


Recent discovery of BCS pairing
in fermionic vapors~\cite{Regal04,Zwierlein04}, made possible by 
control of interactions in trapped
cold gases~\cite{Timmermans99}, has renewed interest in
quantum collective phenomena~\cite{Stringari99}.
Advanced detection techniques and long coherence
times in vapors enable time-resolved studies of
new collective modes, such as
spin waves~\cite{Lewandowski02} and the BCS pairing mode~\cite{Chin04}.

Interaction between a collective mode
and constituting particles is key for our understanding
of dynamics in various systems, from plasma to quantum gases.
One of the most surprising of these phenomena is Landau damping which occurs in
a collisionless regime via direct dissipationless energy transfer from the
collective mode to single particles. Its nondissipative and thus reversible
character~\cite{ONeil65} leads to a variety of regimes, notably
to quenching of the damping, first explored in plasma
physics~\cite{Bernstein57}. Remarkably,
a linearly damped mode can regrow and
transform to a stationary oscillatory
Bernstein-Greene-Kruskal mode.
This fascinating prediction was confirmed experimentally only
recently~\cite{Danielson04}.

Naturally, the richness of these nonlinear phenomena
makes it tempting to look for their analogs in cold gases.
Collisionless damping in cold gases was considered, in the linear regime,
for optical excitations~\cite{Oktel99}, spin waves~\cite{Nikuni03,Ragan05},
and excitations in optical lattices~\cite{Tsuchiya05}.
Motivated by the work on fermion
superfluidity~\cite{Regal04,Zwierlein04,Chin04},
here we focus on the pairing dynamics 
of fermions~\cite{Barankov04-1,Andreev04,Szymanska05,Warner05} 
induced by a sudden change of interaction.
The collisionless regime
becomes practical
in this case 
due to long elastic collision times
$\tau_{\rm el}\gg\tau_\Delta=\hbar/\Delta$~\cite{Barankov04-1},
where $\Delta$ is the BCS gap.
The pairing mode of a small amplitude 
oscillates at a frequency $2\Delta/\hbar$ and
exhibits collisionless dephasing~\cite{Volkov74}.
These conclusions
were extended
recently to the nonlinear regime~\cite{Yuzbashyan05-2}.

This behavior changes drastically as the perturbation increases.
The main result of this work, as summarized in Fig.\ref{fig:dyn_regimes},
is prediction of a dynamical transition
resulting from competition
between synchronization and collisionless dephasing,
taking place as
a function of the initial pairing gap, $\Delta_s$.
We found three qualitatively different regimes
($A$, $B$, and $C$)
with the critical points at
$\Delta_{AB}=e^{-\pi/2}\Delta_0$ and $\Delta_{BC}=e^{\pi/2}\Delta_0$, where
$\Delta_0$ is the equilibrium pairing amplitude in the final
BCS state.
Below the $A$-$B$ transition, $\Delta_s<\Delta_{AB}$,
individual Cooper pair states synchronize
and the pairing amplitude
oscillates between $\Delta_+$ and $\Delta_-$ without damping.
In contrast, in the interval $\Delta_{AB}<\Delta_s<\Delta_{BC}$
the pairing amplitude is Landau-damped and exhibits decaying oscilation,
saturating at an asymptotic
value, $\Delta_a$, with non-monotonic dependence on $\Delta_s$.
A second transition occurs at
$\Delta_s\ge\Delta_{BC}$, where the dynamics
becomes overdamped, and $\Delta(t)$
decreases to zero without oscillations.
The oscillation amplitude and the asymptotic value $\Delta_a$
vanish continuously at the critical points $A$-$B$ and $B$-$C$,
as in a type II transition.
We demonstrate that these results are consistent with the spectral analysis
\cite{Yuzbashyan05} based on the integrability of the problem.

We also address the behavior on a long time scale, $t\gtrsim
\tau_{\rm el}$, after dissipation sets in.
We find that energy relaxation in a closed system, such as an atom trap,
makes it evolve to a new equilibrium state.
Both the temperature $T_*$ and the gap $\Delta(T_*)$
exhibit a nonmonotonic dependence on the initial conditions
(Fig.1).

In our analysis of the BCS problem we employ the well known pseudospin
formulation~\cite{Anderson58} in which spin $1/2$ operators
$s^{\pm}_\vec p=s^x_\vec p\pm is^y_\vec p$ describe Cooper pairs
$(\vec p,-\vec p)$. The BCS Hamiltonian takes the form
\be\label{eq:Hspin}
\HH = -\sum_{\vec p} 2\epsilon_{\vec p} s^z_\vec p-\lambda(t) \sum_{\vec p,\vec
q}s^-_\vec p s^+_\vec q,
\ee
where
$\epsilon_\vec p=\vec p^2/2m-\mu$ is the free particle spectrum with $\mu$ the
Fermi energy. Here we consider the time evolution induced by
an instantaneous change of
interaction from $\lambda_s$ at $t<0$ to $\lambda$ at $t>0$.
%
%
%
%
In the spin formulation, Eq.(\ref{eq:Hspin}), the dynamics is of a Bloch form
%
\be\label{eq:Bloch}
\frac{d\vec r_\vec p}{dt}=2\vec b_\vec p\times \vec r_\vec p,
\quad
\vec b_\vec p=-(\Delta_x,\Delta_y,\epsilon_\vec p),
\ee
where $\vec r_\vec p=2\la \vec s_\vec p\ra$ are classical vectors, and
the effective magnetic field $\vec b_\vec p$
depends on the pairing amplitude $\Delta$.
The latter is defined self-consistently:
\be\label{eq:Delta_def}
\Delta=\Delta_x+i\Delta_y=\frac{\lambda(t)}{2}\sum\limits_{\vec p}r^+_\vec p,
\quad
r^+_\vec p=r^x_\vec p+i r^y_\vec p
.
\ee
%
We first present numerical results for the dynamics
(\ref{eq:Bloch}), (\ref{eq:Delta_def}).
The Runge-Kutta method of the 4th order was used
with $N=10^4,10^5$ equally spaced discrete energy states within a band
$W= 50\Delta_0$ with
a constant density of states $\nu(E_F)$.
As an initial state we take
\be\label{eq:BCS_initial}
r_\vec p^+(0)=\frac{\Delta_{s}}{\sqrt{\Delta_{s}^2+\epsilon_\vec p^2}},\quad
r_\vec p^z(0)= \frac{\epsilon_\vec p}{\sqrt{\Delta_{s}^2+\epsilon_\vec p^2}}
.
\ee
which describes the $T=0$ paired ground
state~\cite{Anderson58}.
Without loss of generality we set $\Delta(t)=\Delta_x$,
since the phase of $\Delta$ is a constant of motion
due to the
particle-hole symmetry of the model.
The interaction constants $\lambda_s$ and $\lambda$ define,
via the self-consistency relation (\ref{eq:Delta_def}),
the initial and final equilibrium BCS gap values,
$\Delta_s=We^{-1/g_s}$, $g_s=\lambda_s\nu\ll 1$,
$\Delta_0=We^{-1/g}$, $g=\lambda\nu$,
which we use to parameterize the system.

\begin{figure}[t]
\includegraphics[width=3.4in]{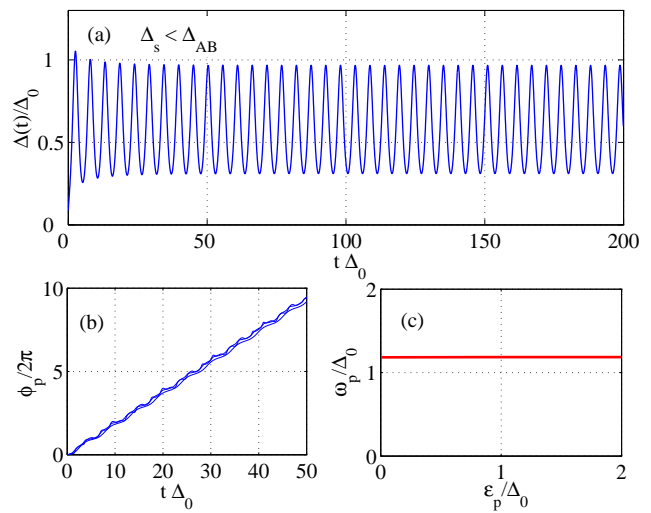}
\vspace{-0.25cm}
\caption[]{(a): The pairing amplitude $\Delta(t)$ for the
initial state~(\ref{eq:BCS_initial}) with $\Delta_s=0.05\Delta_0$ as recorded
from the simulation, oscillating between $\Delta_+=0.97\Delta_0$ and
$\Delta_-=0.31\Delta_0$;
Synchronization (b,c):
the phase $\phi_\vec p$ time dependence, Eq.(\ref{eq:phase}),
for $\epsilon_\vec p=0,
\Delta_0,2\Delta_0$, and frequency $\omega_\vec p$ {\it vs.} $\epsilon_\vec p$,
Eq.(\ref{eq:phase}).
}
\label{fig:DS005D0}
\vspace{-5mm}
\end{figure}

\begin{figure}[t]
\includegraphics[width=3.4in]{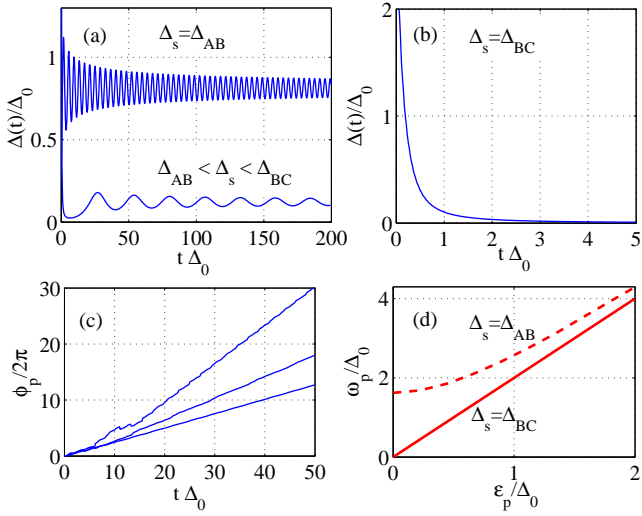}
\vspace{-0.25cm}
\caption[]{Dephased dynamics (a): Simulated $\Delta(t)$ for the
initial states~(\ref{eq:BCS_initial}) with $\Delta_s=0.21\Delta_0,\,
4.5\Delta_0$
with the asymptotic values $\Delta_a=0.81\Delta_0,\, 0.12\Delta_0$;
Overdamped dynamics (b):
Same as
in (a) with $\Delta_s=4.81\Delta_0$ and $\Delta_a=0$;
(c): The
phase $\phi_\vec p$
time dependence,
Eq.(\ref{eq:phase}), for energies
$\epsilon_\vec p=0,\Delta_0,2\Delta_0$ (bottom to top) for
$\Delta_s=0.21\Delta_0$; (d): The frequency $\omega_\vec p$ {\it vs.}
$\epsilon_\vec p$ for $\Delta_s=0.21\Delta_0$ (dashed line) and
$\Delta_s=4.81\Delta_0$ (solid line).}
\label{fig:DS02D0}
\vspace{-3mm}
\end{figure}

We observe three qualitatively different dynamical regimes. The
initial states with a relatively small gap
give rise to undamped oscillations (Fig.\ref{fig:DS02D0}a).
In this case $\Delta(t)$ oscillates non-harmonically between $\Delta_-$
and $\Delta_+$ (the regime $A$ in Fig.\ref{fig:dyn_regimes}).
Synchronization of different Cooper pair states results from
their interaction with the mode singled out by BCS instability
of the initial state, similar to the evolution from the normal
state~\cite{Barankov04-1}.

Desynchronization takes place at $\Delta_s\ge\Delta_{AB}=0.21\Delta_0$
giving rise to two different regimes exhibiting dephasing,
underdamped and overdamped ($B$ and $C$, Fig.~\ref{fig:dyn_regimes}).
The former, illustrated in Fig.\ref{fig:DS02D0}a,
is simplest to understand for a small initial deviation,
$\Delta_s\simeq
\Delta_0$~\cite{Volkov74}, by linearizing Bloch equations about the equilibrium
state. The analysis predicts damped oscillations
at long times:
\be\label{eq:Gap_asym}
\Delta(t)=\Delta_a+A(t)\sin(2\Delta_a t+\alpha),
\quad
A(t)\propto t^{-1/2}.
\ee
The power-law decay of $A(t)$ was explained in Ref.~\cite{Volkov74}
by interaction of the collective mode with the continuous spectrum of
excitations with energies above $2\Delta_a$ and linked
to the linear Landau damping. In the spin formulation,
the dephasing results from the Larmor frequency of spin precession 
$\vec b_\vec p$ being a continuous function
of $\epsilon_\vec p$.
An extension of this argument
to the nonlinear regime was proposed recently
in Ref.~\cite{Yuzbashyan05-2}
which, however, did not clarify the range of its validity.
The dephased time evolution similar to (\ref{eq:Gap_asym}) was also
reported in Refs.~\cite{Szymanska05,Warner05,Amin04}.

In the overdamped regime,
$\Delta_s\ge\Delta_{BC}=4.81\Delta_0$ (Fig.\ref{fig:DS02D0}b),
$\Delta(t)$ decays to zero without oscillations.
This behavior can be understood in the limit
$\Delta_s/\Delta_0\gg1$, i.e. when
the coupling in the initial paired state in (\ref{eq:Hspin})
is suddenly completely turned off.
For different spins precessing freely and independently
one obtains
%
\[
r^+_\vec p(t)=e^{-i2\epsilon_\vec p t}r_\vec p^+(0),
\quad
\Delta(t\gg\Delta_s^{-1})\propto
(\Delta_s t)^{-1/2}e^{-2\Delta_s t}.
%
\]
%
The fast dephasing can also be understood by noting that the energy
distribution in (\ref{eq:BCS_initial}) corresponds to
an effective temperature $T\sim \Delta_s$
which exceeds $T_c$ for $\Delta_0$ (see below).



To fully exhibit phase locking in the synchronized
regime which abruptly disappears in the dephased regime,
we now explore the phase dynamics. It is convenient to
measure precession angles relative to time-independent
$\tilde\vec b_\vec p=-(\Delta_a,0,\epsilon_\vec p)$, where
$\Delta_a$ is the asymptote $\Delta(t\to\infty)$
in the regimes $B$, $C$, and the
average value of oscillating $\Delta(t)$ in $A$
(dash-dotted line in Fig.~\ref{fig:dyn_regimes}).
The angle and frequency of precession
are defined by
\be\label{eq:phase}
n^+_\vec p=n^x_\vec p+in^y_\vec p\propto e^{-i\phi_\vec p(t)}
,\quad
\omega_\vec p(t)=d\phi_\vec p/dt
,
\ee
where the vectors $n_\vec p$ are obtained from $r_\vec p$
by a rotation about the $y$ axis which maps
$\hat z$ onto $\tilde\vec b_\vec p$:
$n^y_\vec p=r^y_\vec p$, $n^x_\vec p+in^z_\vec p=e^{i\theta_\vec
p}(r^x_\vec p+ir^z_\vec p)$,
with the rotation angle defined by
$\tan\theta_\vec p=\Delta_a/|\epsilon_\vec p|$.
The phase evolution,
which becomes linear at long times $\tau\gg\Delta_0^{-1}$
(Figs.\ref{fig:DS005D0},\ref{fig:DS02D0}),
can be characterized by average frequency (phase slope)
$\omega_\vec p = \la d\phi_\vec p/dt\ra=(\phi_\vec
p(\tau)-\phi_\vec p(0))/\tau$.
While in the regime $A$ different $\vec p$ states phase lock
(Fig.\ref{fig:DS005D0}b,c), in the regimes $B$, $C$ the frequencies
$\omega_\vec p$ have dispersion (Fig.\ref{fig:DS02D0}c,d).
The latter reproduces quasiparticle spectrum,
$\omega_\vec p=2(\epsilon_\vec p^2+\Delta_a^2)^{1/2}$
with the long-time asymptote $\Delta_a$ which vanishes in the
overdamped regime $C$.

We observe a qualitative change in behavior,
with $\omega_\vec p$ dispersing
in the regions $B$, $C$ and phase locking in the region $A$.
However, the oscillation
amplitude $\frac12(\Delta_+-\Delta_-)$ in $A$
and the asymptotic amplitude $\Delta_a$ in $B$
vanish continuously at the critical points
$\Delta=\Delta_{AB},\Delta_{BC}$
(Fig.\ref{fig:dyn_regimes}),
indicating a type II transition.

Until now, we considered the dissipationless dynamics at times shorter than the
quasiparticle relaxation time, $t\lesssim \tau_{\rm el}$. Using
the energy balance argument, one can determine the
system state
at long times, $t\gg \tau_{\rm el}$.
To account for system equilibration,
one needs to consider the full
many-body Hamiltonian which enables elastic scattering of
individual quasiparticles, omitted in Eq.(\ref{eq:Hspin}).
For a closed system, such as an atomic trap,
the final temperature $T_*$ and the gap $\Delta_*$ can be
determined from the energy conservation  condition
$E_{t\gg\tau_{\rm el}}(T_*)=E_{0}$,
%
%
%
%
where $E_{0}$ is the energy immediately after interaction switching,
\be
E_{0}=\sum_\vec p \lp\epsilon_\vec p-\tilde\epsilon_\vec p\rp
+\frac{\Delta_s^2}{\lambda_s}\lp 2-\frac{\lambda}{\lambda_s}\rp,
\ee
with the spectrum $\tilde\epsilon_\vec p=(\epsilon_\vec p^2+\Delta_s^2)^{1/2}$,
and $E_{t\gg\tau_{\rm el}}(T_*)$ is the energy of the final state:
%
\be
E_{t\gg\tau_{\rm el}}(T_*)=\sum_\vec p\lb\epsilon_\vec p-\lp 1- 2n_\vec p\rp
\tilde\epsilon_\vec p(\Delta_*)\rb +\frac{\Delta^2_*}{\lambda}.
\ee
Here $n_\vec p=1/\lp 1+e^{\tilde\epsilon_\vec p(\Delta_*)/T_*}\rp$
describes equilibrium with $T=T_*$,
$\tilde\epsilon_\vec p(\Delta_*)=(\epsilon_\vec p^2+\Delta^2_*)^{1/2}$.
After integrating over $\epsilon_\vec p$, we arrive at the
equation for $T_*$:
%
%
\be\label{eq:T_final}
F\lp\frac{\Delta_*}{2T_*}\rp=
1-\lp\frac{\Delta_s}{\Delta_*}\rp^2+ \alpha \lp\frac{\Delta_s}{\Delta_*}\rp^2
\ln\lp\frac{\Delta_s}{\Delta_0}\rp^2,
%
\ee
where $F(u)=2\int_0^\infty dx \cosh 2x \lb 1-\tanh\lp u\cosh x\rp\rb$, and
$\alpha=  1-g\ln(\Delta_s/\Delta_0)$.
From Eq.(\ref{eq:T_final}) we obtain $T_*$ and the equilibrium gap
$\Delta_*=\Delta(T_*)$. They are approximately constant,
with $T_*\approx 0.72 T_c$, $\Delta(T_*)\approx 0.81\Delta_0$
in the regime $A$, are described by
a non-monotonic function in $B$, and vanish in $C$.
The system turns normal in the final state 
for $\Delta_s\ge f(g)\Delta_0$,
$f(g)= 2.2+0.86g+O(g^2)$.
Numerical solution of Eq.(\ref{eq:T_final}) at
$g=0.26$ is displayed in Fig.\ref{fig:dyn_regimes} (dashed line).

\begin{figure}[t]
\includegraphics[width=3.4in]{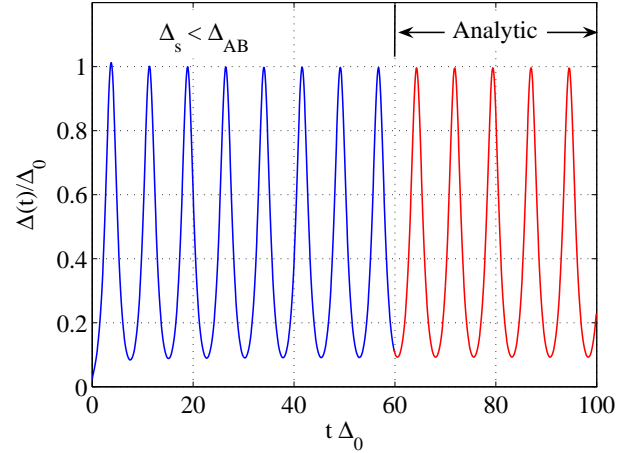}
\vspace{-0.25cm}
\caption[]{The pairing amplitude $\Delta(t)$ as a function of time for the
initial state~(\ref{eq:BCS_initial}) with $\Delta_s=0.01\Delta_0$ as recorded
from numerical simulation. At long time, numerical $\Delta(t)$ matches the
analytic form in Eq.(\ref{eq:DN}), oscillating non-harmonically between
$\Delta_+=0.997\Delta_0$ and $\Delta_-=0.093\Delta_0$.}
\label{fig:DS001D0}
\vspace{-3mm}
\end{figure}

It is instructive to compare our results
to the spectral analysis based on the integrability
of the BCS Hamiltonian~\cite{Richardson64,Cambiaggio97,Yuzbashyan05}.
There is an
infinite number of commuting integrals of motion, $R_\vec p=\vec L_\vec p\vec
s_\vec p$,
parameterized by $\vec p$,
%
%
where following Ref.~\cite{Yuzbashyan05} we employ the Lax vector,
\be\label{L_p}
\vec L_\vec p=\hat z+\lambda\sum_{\vec p'\ne \vec p}\frac{\vec s_{\vec
p'}}{\epsilon_\vec p-\epsilon_{\vec p'}}.
\ee
We will need $\vec L_\vec p^2=4(\vec L_\vec p\vec s_\vec p)^2$
which is also conserved (due to the Pauli matrices
algebra combined with Eq.(\ref{L_p})).
The mean-field
expressions are obtained by substituting the
averages $\la\vec s_\vec p\ra$ instead
of $\vec s_\vec p$. 

The spectral polynomial defining the evolution
of individual states is proportional to the square
of the Lax vector~\cite{Yuzbashyan05},
$Q(\epsilon)=\vec L^2(\epsilon)\prod_\vec p(\epsilon-\epsilon_\vec p)^2$, where
$\vec L(\epsilon_\vec p)=\vec L_\vec p$. The pairs of complex roots of the
spectral equation $Q(\epsilon)=0$ uniquely determine the long-time dynamics of
the system~\cite{Yuzbashyan05}. Evaluation of
$\vec L^2(y)$ for
the initial state~(\ref{eq:BCS_initial}) is straightforward:
\be
\vec L^2(y)/g^2=\lp\ln\frac{\Delta_s}{\Delta_0}+yG(y)\rp^2+G^2(y),
\ee
%
where $y=\epsilon/\Delta_s$ and
\[
G(y)=\frac12\int_{-\infty}^{\infty}\frac{dx}{y-\sinh x}
=\frac{1}{2\sqrt{1+y^2}}\ln\frac{y+\sqrt{1+y^2}}{y-\sqrt{1+y^2}}
,
\]
where $y$ is complex.
%
%
%
%
%
%
One can show that the roots of $\vec L^2(y)$ 
lie on the imaginary axis. Upon variation of $\Delta_s/\Delta_0$
they disappear at $y=0$.
Expanding about $y=0$, we obtain
%
\be\label{eq:Lax_zeros}
\vec L^2(y)/g^2= \lp\ln\frac{\Delta_s}{\Delta_0}-\frac{\pi}{2}\rp
\lp\ln\frac{\Delta_s}{\Delta_0}+\frac{\pi}{2}\rp+O(y),
\ee
%
Thus $\vec L^2(y)$ has no complex roots at
$\Delta_s/\Delta_0\ge e^{\pi/2}$,
one pair of roots
$y=\pm iu$
when $e^{-\pi/2}\le\Delta_s/\Delta_0\le e^{\pi/2}$, with another pair appearing
at
$\Delta_s/\Delta_0\le e^{-\pi/2}$.

There is a direct correspondence between this behavior
of the roots and
the dynamical regimes $A$, $B$, and $C$ observed numerically.
The pairing amplitude is subject to fast dephasing and tends to zero when
$\vec L^2(y)$ does not have
complex roots. A pair of complex roots $y_a=\pm
i\Delta_a/\Delta_s$ defines the long-time asymptote
$\Delta(t)\approx \Delta_a$.
Two pairs of roots, $y=\pm i u_1$ and $y=\pm i u_2$,
correspond to the parameters
$\Delta_\pm =(u_1\pm u_2)\Delta_s$
of the elliptic
function which defines the asymptotic behavior:
%
\be\label{eq:DN}
\Delta(t)=\Delta_+{\rm dn}\lb\Delta_+(t-\tau_0),k\rb
,\quad
k=1-\Delta_-^2/\Delta_+^2
,
\ee
%
%
%
where the time lag $\tau_0$ is a half of the period. As illustrated in
Fig.\ref{fig:DS001D0},
Eq.(\ref{eq:DN}) agrees well with $\Delta(t)$ found numerically.
Thus the spectral analysis is in accord with
the simulation of Bloch
dynamics.
It confirms the existence of the three regimes and also provides
the exact values
$\Delta_{AB}=e^{-\pi/2}\Delta_0$ and
$\Delta_{BC}=e^{\pi/2}\Delta_0$.

Finally, we estimate the change of scattering length
required to cross the $A$-$B$ and $B$-$C$ transitions.
Using the BCS gap in a weakly interacting Fermi gas\cite{Gorkov61},
$\Delta=0.49 E_F e^{-1/g}$, $g=\frac{2}{\pi}k_F|a|$,
we see that the conditions $\Delta_s/\Delta_0=e^{\pm\pi/2}$,
written as  $1/g-1/g_s=\pm\pi/2$,
translate
into $1/a-1/a_s=\pm k_F$. At weak coupling this corresponds to a small
change of scattering length, $\delta a/a\approx\pm k_F a$,
easily achievable
for magnetically tunable Feshbach
resonance.

We are grateful to Boris Spivak and Kumar Raman for helpful discussions.



\begin{thebibliography}{10}

\bibitem{Regal04}
C. A. Regal, M. Greiner, and D. S. Jin, Phys. Rev. Lett. {\bf 92}, 040403
(2004).

\bibitem{Zwierlein04}M. W. Zwierlein, C. A. Stan, C. H. Schunck, S. M. F. Raupach,
A. J. Kerman, and W. Ketterle,
Phys. Rev. Lett. {\bf 92}, 120403 (2004).

\bibitem{Timmermans99} E. Timmermans, P. Tommasini, M. Hussein,
and A. Kerman, Physics Reports {\bf 315}, 199 (1999).

\bibitem{Stringari99}
F. Dalfovo, S. Giorgini, L. P. Pitaesvkii, and S. Stringari, Rev. Mod. Phys.
{\bf  71}, 463 (1999).

\bibitem{Lewandowski02}
H. J. Lewandowski, D. M. Harber, D. L. Whitaker,
and E. A. Cornell, Phys. Rev. Lett. {\bf 88}, 070403 (2002);
%
J. M. McGuirk, H. J. Lewandowski, D. M. Harber, T. Nikuni, J. E.
Williams, and E. A. Cornell, Phys. Rev. Lett. {\bf 89}, 090402 (2002).

\bibitem{Chin04}
C. Chin, M. Bartenstein, A. Altmeyer, S. Riedl, S. Jochim, J. H. Denschlag, and
R. Grimm, Science, {\bf 305}, 1128 (2004).

\bibitem{ONeil65}
T. M. O'Neil, Phys. Fluids {\bf 8}, 2255 (1965).

\bibitem{Bernstein57}
I. B. Bernstein, J. M. Green, and M. D. Kruskal,
Phys. Rev. {\bf 108}, 546 (1957)

\bibitem{Danielson04}
J. R. Danielson, F. Anderegg, and C. F. Driscoll,
Phys. Rev. Lett. {\bf 92}, 245003 (2004).
















%

\bibitem{Oktel99}
M.\"O. Oktel and L. S. Levitov, Phys. Rev. Lett. {\bf 83}, 6 (1999).

\bibitem{Nikuni03} T. Nikuni and J. E. Williams, J. Low Temp. Phys. {\bf 133}, 323 (2003).


\bibitem{Ragan05} R. J. Ragan, W. J. Mullin, and E. B. Wiita, cond-mat/0502189

\bibitem{Tsuchiya05} S. Tsuchiya and A. Griffin, Phys. Rev. A {\bf 72}, 053621 (2005).



\bibitem{Barankov04-1}R. A. Barankov, L. S. Levitov, and B. Z. Spivak,
Phys. Rev. Lett. {\bf 93}, 160401 (2004).

\bibitem{Andreev04}
A. V. Andreev, V. Gurarie, and L. Radzihovsky, Phys. Rev. Lett. {\bf 93},
130402 (2004); R. A. Barankov and L. S. Levitov, Phys. Rev. Lett. {\bf 93},
130403 (2004).

\bibitem{Szymanska05}M. H. Szyma\'nska, B. D. Simons, and K. Burnett,
Phys. Rev. Lett. {\bf 94}, 170402 (2005).

\bibitem{Warner05}G. L. Warner and A. J. Leggett, Phys. Rev. B {\bf 71}, 134514
(2005).

\bibitem{Volkov74} A. F. Volkov and Sh. M. Kogan, Zh. Eksp. Teor. Fiz.
{\bf 65}, 2038, (1973) [Sov. Phys. JETP {\bf 38}, 1018 (1974)].

\bibitem{Yuzbashyan05-2}E. A. Yuzbashyan, O. Tsyplyatyev, and B. L. Altshuler,
Phys. Rev. Lett. {\bf 96}, 097005 (2006).

\bibitem{Yuzbashyan05}E. A. Yuzbashyan, B. L. Altshuler, V. B. Kuznetsov,
and V. Z. Enolskii, Phys. Rev. B {\bf 72}, 220503 (2005).

\bibitem{Anderson58}P. W. Anderson, Phys. Rev. {\bf 112}, 1900 (1958)



\bibitem{Amin04}
M. H. S. Amin, E. V. Bezuglyi, A. S. Kijko, and A. N. Omelyanchouk, Low Temp.
Phys. {\bf 30}, 661 (2004), cond-mat/0404401

\bibitem{Richardson64}R. W. Richardson and N. Sherman, Nucl. Phys. {\bf 52},
221 (1964).

\bibitem{Cambiaggio97}M. C. Cambiaggio, A. M. F. Rivas, and M. Saraceno, Nucl.
Phys. A {\bf 424}, 157 (1997).


\bibitem{Gorkov61} L. P. Gor'kov and T. K. Melik-Barkhudarov,
Zh. Eksp. Teor. Fiz. {\bf 40}, 1452, (1961) [Sov. Phys. JETP, {\bf 13},  1018
(1961)].




\end{thebibliography}
\end{document}